\documentclass[12pt]{iopart}
\usepackage{graphicx,amssymb,url,epstopdf}
\usepackage[normalem]{ulem}
\begin{document}

\title[Access and sustainment of ELMy H-mode for ITER PFPO using JINTRAC]{Access and sustainment of ELMy H-mode operation for ITER Pre-Fusion Power Operation plasmas using JINTRAC}

\author{E Tholerus$^1$, L Garzotti$^1$, V Parail$^1$, Y Baranov$^1$, X Bonnin$^2$, G Corrigan$^1$, F Eriksson$^1$, D Farina$^3$, L Figini$^3$, D M Harting$^4$, S H Kim$^2$, F Koechl$^1$, A Loarte$^2$, E Militello Asp$^1$, H Nordman$^5$, S D Pinches$^2$, A R Polevoi$^2$, P Strand$^5$}

\address{%
$^1$ UKAEA (United Kingdom Atomic Energy Authority), Culham Campus, Abingdon, Oxfordshire, OX14 3DB, UK\\
$^2$ ITER Organization, Route de Vinon-sur-Verdon, CS 90\,046, 13067 St.~Paul Lez Durance Cedex, France\\
$^3$ Istituto per la Scienza e Tecnologia dei Plasmi, CNR, Milan, Italy\\
$^4$ Institut für Energie- und Klimaforschung IEK-4, FZJ, TEC, 52425 Jülich, Germany\\
$^5$ Association EURATOM-VR, Chalmers University of Technology, Göteborg, Sweden}
\ead{emmi.tholerus@ukaea.uk}

\begin{abstract}
In the initial stages of ITER operation, ELM mitigation systems need to be commissioned. This requires controlled flat-top operation in type-I ELMy H-mode regimes. Hydrogen or helium plasma discharges are used exclusively in these stages to ensure negligible production of neutrons from fusion reactions. With the expected higher L--H power threshold of hydrogen and helium plasmas compared to corresponding D and D/T plasmas, it is uncertain whether available auxiliary power systems are sufficient to operate in stable type-I ELMy H-mode. This has been investigated using integrated core and edge/SOL/divertor modelling with JINTRAC. Assuming that the L--H power threshold is well captured by the Martin08 scaling law, the presented simulations have found that 30\,MW of ECRH power is likely required for the investigated hydrogen plasma scenarios, rather than the originally planned 20\,MW in the 2016 Staged Approach ITER Baseline. However, past experiments have shown that a small helium fraction ($\sim$10\,\%) can considerably reduce the hydrogen plasma L--H power threshold. Assuming that these results extrapolate to ITER operation regimes, the 7.5MA/2.65T hydrogen plasma scenario is likely to access stable type-I ELMy H-mode operation also at 20\,MW of ECRH.
\end{abstract}
%
\vspace{2pc}

\submitto{\NF}
%
%
%

\section{Introduction}

The ITER Pre-Fusion Power Operation (PFPO) is an important step in the ITER research plan~\cite{IRP} to demonstrate full technical capability of the ITER tokamak and to prepare for the main D and D--T campaigns. It is intended to demonstrate stable H-mode operation, as well as to commission several systems, such as auxiliary heating and current drive, fuelling, various diagnostics, ELM mitigation, and divertor heat flux control. PFPO will consist of two sub-phases. The first one, PFPO-1, will operate with at least 20\,MW ECRH heating and current drive, and a selected set of diagnostics, fuelling and scenario control capabilities. The second sub-phase, PFPO-2, will operate with the full set of resources planned for baseline operation, including the ECRH capability of PFPO-1, 33\,MW of hydrogen NBI and 20\,MW ICRF auxiliary heating and current drive.\footnote{Following significant delays of the ITER experimental programme, an updated research plan is being formulated at the time of writing this paper. The updated research plan will likely impact the selection of scenarios during PFPO, the available auxiliary heating systems in different stages of ITER operation, and the wall material composition. All work presented here is based on the assumptions of the 2018 research plan~\cite{IRP}.} All plasmas during PFPO will have hydrogen or helium as main ion species to ensure non-active operation.

This paper presents modelling of PFPO scenarios mainly intended for com\-mis\-sion\-ing of the ELM mitigation systems, which requires stable operation in type-I ELMy H-mode. The presented scenarios are 5MA/1.8T and 7.5MA/2.65T, both operating with hydrogen as main ion species. Whether 20\,MW ECRH is sufficient for stable ELMy H-mode operation or an additional 10\,MW ECRH is required will be investigated. Methods for lowering the L--H power threshold will also be considered. All simulations are performed with integrated core, edge and SOL/divertor modelling using JINTRAC~\cite{jintrac}, developed by EUROfusion. Flat-top stages of the scenarios are presented here, starting from L-mode and setting up the density and auxiliary power required for triggering an L--H transition. The H-mode is sustained until the edge ballooning parameter $\alpha$ stabilizes, which is then compared against the critical level for type-I ELM destabilisation, $\alpha_\mathrm{crit}$.

\section{Modelling assumptions}\label{model}

\subsection{Integrated modelling structure}

\begin{figure}\centering
\includegraphics[width=\textwidth]{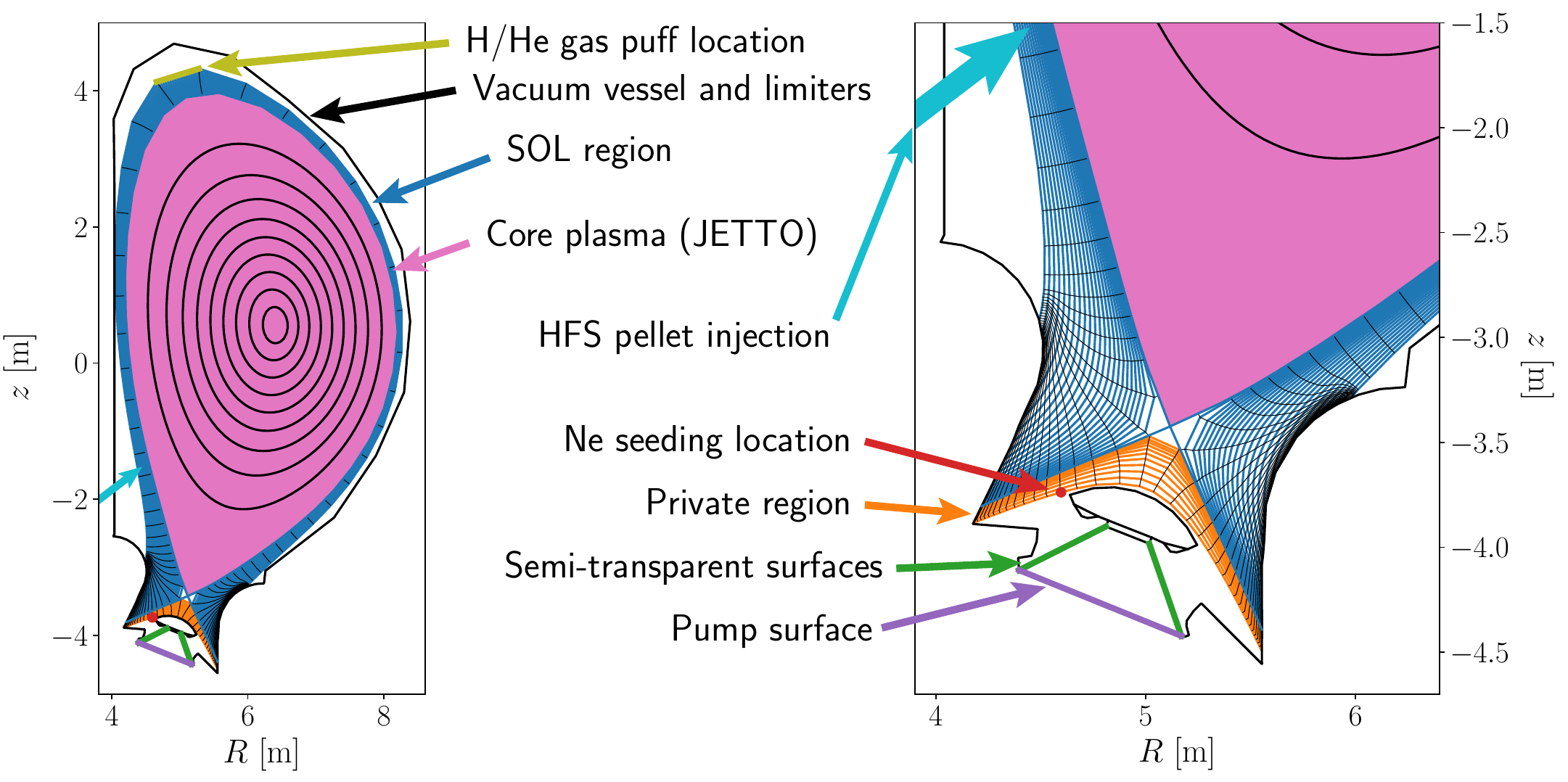}
\caption{2D geometric set-up for the presented JINTRAC simulations.}\label{fig:jin}
\end{figure}

\begin{figure}\centering
\includegraphics[width=\textwidth]{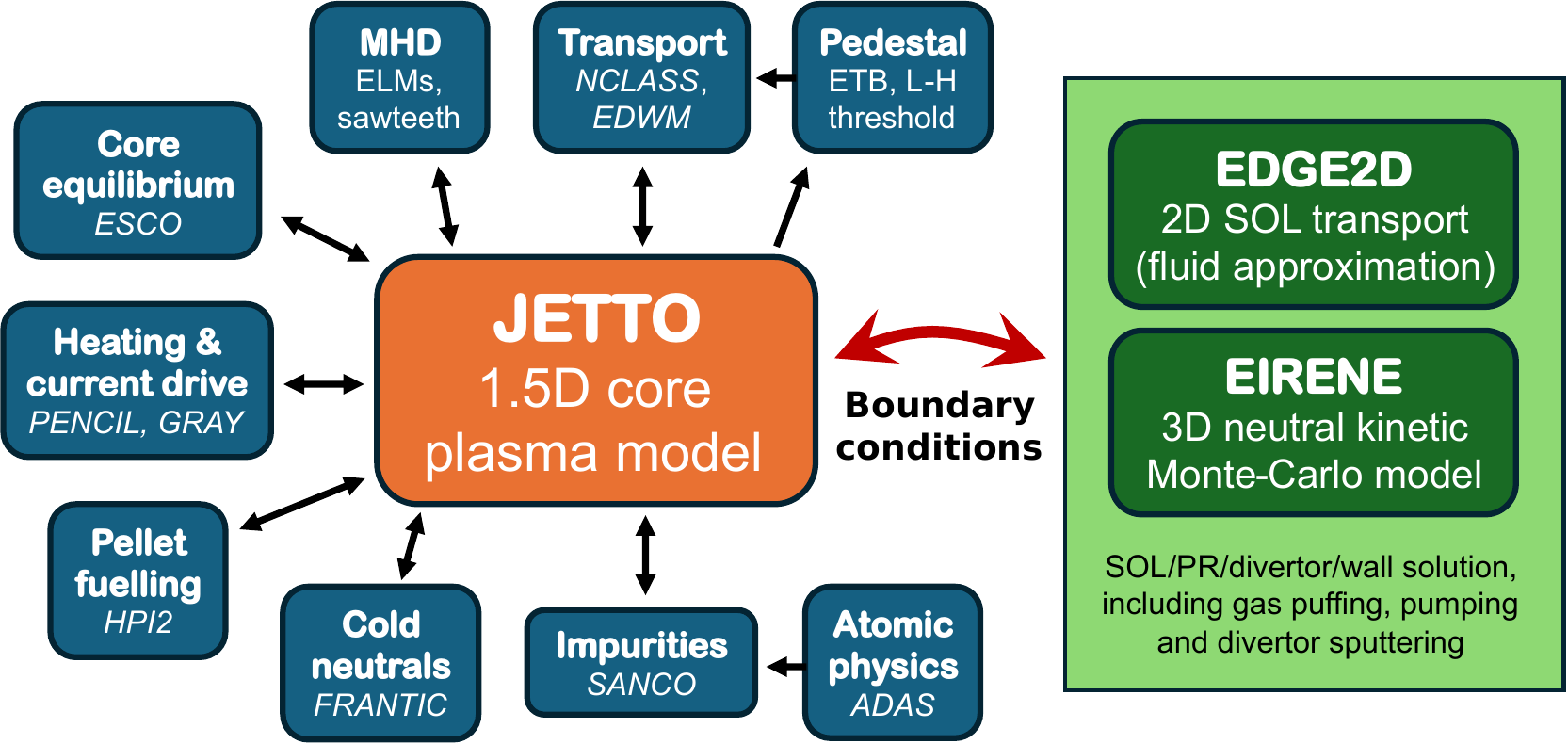}
\caption{Flowchart for JINTRAC integrated modelling. In italics are different codes/databases integrated with JINTRAC, whereas models for e.g. MHD, edge transport barrier (ETB) and L-H power threshold are intrinsic to JETTO.}\label{fig:flo}
\end{figure}

JINTRAC~\cite{jintrac} is an integrated tokamak plasma model for simulation of the whole plasma, including core, pedestal, scrape-off layer (SOL), and private region (PR). It includes models for heating and current drive, self-consistent equilibrium calculation, transport (neoclassical and anomalous), fuelling (gas puffing and pellet injection), pumping, sputtering, atomic physics, sawteeth and ELMs. The two main components of JINTRAC are JETTO~\cite{jetto} and EDGE2D/EIRENE~\cite{e2d1,e2d2,e2d3}. JETTO is a 1.5-dimensional core plasma model that simulates the plasma inside the last closed flux surface, whereas EDGE2D is a 2D SOL/PR Braginskii fluid model, and EIRENE simulates neutrals in the SOL/PR, including models for gas puffing, pumping, sputtering and recycling. EDGE2D/EIRENE is generally more computationally demanding than JETTO for a given interval of plasma time. For numerical efficiency, JINTRAC typically utilizes a partial coupling scheme between the codes, where JETTO and EDGE2D/EIRENE are fully coupled only during short intervals. In between the coupled phases, JETTO runs stand-alone for a selected interval (usually 2 -- 20 times the coupled time interval). Fluxes are then rescaled at the EDGE2D/EIRENE boundary to compensate for the inactive time interval.

Figure \ref{fig:jin} presents the geometric assumptions of the modelling, including first wall and divertor, fuelling and pumping surface locations. The EDGE2D grid, shown in blue and orange in the figure, is fixed, effectively fixing the geometry of the separatrix. Figure \ref{fig:flo} shows the workflow between different parts of JINTRAC. Boundary conditions between JETTO and EDGE2D/EIRENE are applied at the last closed flux surface. The following subsections describe how each component of JINTRAC is set-up in more detail.

\subsection{Heating and current drive}

The 7.5MA/2.65T hydrogen plasma scenario is planned for PFPO-2, and is heated by a combination of ECRH and NBI. The 5MA/1.8T scenario is part of PFPO-1, and consequently does not include heating by NBI. Neither of the presented cases include heating by ICRF. Heating by conventional ICRF schemes is not efficient for the 7.5MA/2.65T hydrogen scenario~\cite{IRP,lerche}, whereas the 5MA/1.8T first H-mode scenario in the PFPO-1 phase has no installed ICRF power. However, there is a possibility that efficient heating based on three-ion schemes exists for the hydrogen scenarions with He-3 and He-4 at 8.8MA/3.13T~\cite{icrh,kaz}. Two different ECRH power schemes are considered. The first scheme is 20\,MW using the present baseline design of the ITER ECRH systems. In the second scheme, 10\,MW of ECRH power is added to the baseline, assuming an upgrade of the ITER heating systems that is being assessed~\cite{IRP}. 

ECRH/ECCD is modelled predictively using the GRAY model~\cite{gray}. The equatorial EC launchers operate in O-mode in the 7.5MA/2.65T scenarios, which gives less parasitic absorption and more efficient ECCD compared to X-mode operation during H-mode confinement, and X-mode in the 5MA/1.8T scenario. Some of the presented cases use the upper launcher for the suggested 10\,MW ECRH upgrade, which operates in X-mode in scenarios where it is active. NBI heating and current drive is modelled with PENCIL~\cite{pencil}, operating at full power (16.5\,MW on each of the two negative ion source injectors), injecting hydrogen at $\sim$870\,keV. 

The bootstrap current is evaluated from neoclassical theory with NCLASS~\cite{nclass}. The total plasma current is set as a boundary condition in the simulations, with the inductive current and the corresponding loop voltage being adapted such that it completes the bootstrap and auxiliary currents to reach the set target value. With only flat-top stages being modelled, the total current is set to a constant level of either 5\,MA or 7.5\,MA depending on the scenario. The self-consistent handling of inductive current drive profiles and loop voltage together with current diffusion is described in \cite{jetto}. 

\subsection{L--H transition}\label{sec:PLH}

The assumed L--H power threshold scaling law is based on Martin08~\cite{martin}, with a 
correction $P_\mathrm{L-H} = P_{\mathrm{L-H},\mathrm{Martin08}}\times(2/A_\mathrm{eff})$ for hydrogenic plasmas~\cite{righi}:
\begin{equation}\label{eq:plh}
	P_\mathrm{L-H} = 0.0488\langle n_\mathrm{e,20}\rangle_\mathrm{line}^{0.717}B_\mathrm{tor}^{0.803}S^{0.941}(2/A_\mathrm{eff}),
\end{equation}
where line-averaged electron density $\langle n_\mathrm{e,20}\rangle_\mathrm{line}$ has the unit of 10$^{20}$\,m$^{-3}$, the toroidal magnetic field $B_\mathrm{tor}$ is in T, the cross sectional area of the last closed flux surface $S$ is in m$^2$, and
\begin{equation}
	A_\mathrm{eff} = \frac{\langle n_\mathrm{H} + 2 n_\mathrm{D} + 3 n_\mathrm{T}\rangle}{n_\mathrm{H} + n_\mathrm{D} + n_\mathrm{T}}.
\end{equation}
Since no deuterium or tritium is present in any of the PFPO scenarios, the correction factor $2/A_\mathrm{eff} = 2$. The scaling law of eq.~(\ref{eq:plh}) is expected to have a larger uncertainty at low density, when $\langle n_\mathrm{e}\rangle \lesssim n_{\mathrm{e},\mathrm{min}} \approx 0.4 n_\mathrm{GW}$~\cite{maggi}, corresponding to a density $\langle n_\mathrm{e}\rangle \approx 2.5\times 10^{19}$\,m$^{-3}$ for the 7.5MA scenario, and $\langle n_\mathrm{e}\rangle \approx 1.6\times 10^{19}$\,m$^{-3}$ for the 5MA scenario. The density domain $\langle n_\mathrm{e}\rangle \gtrsim n_\mathrm{e,min}$ where the L--H power threshold follows the above scaling law is commonly referred to as the high-density branch of the L--H transition. 

JET experiments have shown that both $P_\mathrm{L-H}$ and $n_{\mathrm{e},\mathrm{min}}$ are sensitive to the detailed strike-point configuration~\cite{delabie}. It was observed that by moving the outer strike point outwards from the horizontal target (HT) to the vertical target (VT) plate, $n_{\mathrm{e},\mathrm{min}}/n_\mathrm{GW}$ was reduced from around 40\,\% to around 30\,\% when operating JET at 3.0\,T and 2.5 -- 2.75\,MA. The VT configuration also indicated a stronger density scaling than the 0.717 exponent of the Martin08 scaling in the high-density branch. On the other hand, the HT configuration showed similar $P_\mathrm{L-H}$ scaling with respect to density as the Martin08 scaling, but a value roughly 25\,\% lower than $P_\mathrm{L-H,Martin08}$. It was hypothesized that the mechanism behind the dependence of $P_\mathrm{L-H}$ with respect to strike-point configuration is different degree of turbulence suppression by $E\times B$ shear in the edge region. It is non-trivial to translate these dependencies to ITER scenarios due to differences in divertor geometry between JET and ITER. However, the JET results can give an indication of the order of magnitude differences in $P_\mathrm{L-H}$ and $n_{\mathrm{e},\mathrm{min}}$ with respect to strike point configuration. The specific values of $P_\mathrm{L-H}$ and $n_{\mathrm{e},\mathrm{min}}$ assumed in the presented ITER model have resulted from compromises between several aspects of their respective observed dependencies in past experiments, details of which is beyond the scope of this paper.

JET experiments~\cite{hil} have suggested that adding a certain fraction of helium can reduce $P_\mathrm{L-H}$ of hydrogen plasmas. Therefore, we will also consider the cases with and without added helium, assuming a 15\,\% reduction of $P_\mathrm{L-H}$ when $\langle n_\mathrm{He}\rangle \approx 0.1\langle n_\mathrm{e}\rangle$. Above all, to facilitate the access to H-mode, the basic assumption for L--H power threshold is to operate at densities close to $n_{\mathrm{e},\mathrm{min}} \approx 0.4 n_\mathrm{GW}$, as $P_\mathrm{L-H} \sim \langle n_\mathrm{e}\rangle^{0.717}$~\cite{martin}. It should be noted that JET experiments on helium plasmas have shown that $n_{\mathrm{e},\mathrm{min}}$ can range between $0.4 n_\mathrm{GW}$ and $0.7 n_\mathrm{GW}$ depending on the strike-point configuration~\cite{solano}. This observation would need further investigation to determine how it extrapolates to ITER plasmas and strike-point configurations. The results presented here assume a reduction of $P_\mathrm{L-H}$ by a helium minority at $\langle n_\mathrm{e}\rangle \gtrsim 0.4 n_\mathrm{GW} \approx 2.5\times 10^{19}$\,m$^{-3}$ (cases D and E, as presented below).

H-mode access is determined by comparing $P_\mathrm{L-H}$ against $P_\mathrm{net} = P_\mathrm{ECRH} + P_\mathrm{NBI} + P_\mathrm{ohm} - P_\mathrm{rad} - \langle\mathrm{d}W_\mathrm{p}/\mathrm{d}t\rangle$, where $P_\mathrm{rad}$ is the combined impurity radiation and bremsstrahlung from the core plasma (inside the last closed flux surface), and $\langle\mathrm{d}W_\mathrm{p}/\mathrm{d}t\rangle$ is the time derivative of the total stored energy $W_\mathrm{p}$, averaged over a time window of 5 -- 50\,ms for numerical stability. During H-mode ($P_\mathrm{net} \geq P_\mathrm{L-H}$), the anomalous transport of the edge is continuously lowered by a factor
\begin{equation}
	\theta = \exp\left(-\frac{P_\mathrm{net}-P_\mathrm{L-H}}{P_\mathrm{L-H}\Delta_\mathrm{L-H}}\right),
\end{equation}
where $0.05 \leq \Delta_\mathrm{L-H} \leq 0.1$ is a numerical factor for smoothness of the edge transport barrier formation.

\subsection{Fuelling and impurity seeding}\label{sec:fuel}

Fuelling by hydrogen gas puffing alone is sufficient for low density plasmas ($\langle n_\mathrm{e}\rangle \lesssim 2\times 10^{19}$\,m$^{-3}$). However, at high gas fuelling rates there is a risk of detachment due to excessive cooling of the edge plasma. A detached plasma configuration is generally avoided in the presented simulations. Although detachment access can potentially reduce divertor heat loads, conventional detachment techniques might also reduce core plasma density control. For numerical efficiency, EDGE2D/EIRENE does not include the molecular reactions necessary to accurately model detachment. In order to reduce the risk of detachment access, some of the presented scenarios do hydrogen fuelling by a combination of gas puffing and pellet injection (cases D and E, summarized in Sec.~\ref{sec:scen}). The additional pellet injection also allows for access to higher densities, as will be demonstrated for case E. Pellet ablation and deposition is modelled with HPI2~\cite{hpi2}.

In order to avoid unacceptable levels of NBI shine-through power in lower density regimes that can reduce the life expectancy of the NBI shield blocks, a neon minority can be introduced to the core plasma to increase the beam stopping cross section~\cite{singh}. The required concentration of neon depends on the total density of the plasma, with higher densities requiring less neon for sufficient beam stopping. In these simulations, neon gas rates are adapted to reach stabilized total shine-though power below about 1.8\,MW while avoiding full divertor detachment. The simulations also consider critical upper limits associated with excessive impurity radiation, for instance a reduction of the net power flux across the separatrix due to core radiation, limiting the possibility to sustain stable ELMy H-mode operation, or a fully detached plasma by excessive edge/SOL cooling. Neon is also more efficient than H and He at sputtering tungsten from the divertor, potentially adding significantly to the impurity radiation. 

SANCO~\cite{sanco} is used for modelling impurity atomic physics in the core, including radiation and ionisation/recombination, and EIRENE is used in the SOL and private region. Helium and neon atomic data is based on~\cite{adas96}, and tungsten data is based on~\cite{adas41}. Charge state bundling schemes have been used both for neon (5 charge state levels) and tungsten (6 charge state levels), rather than following all ionisation stages for these impurities.

\subsection{Transport}

Core heat and particle transport for both the main plasma and impurity species are handled by NCLASS~\cite{nclass} (neoclassical transport) and EDWM~\cite{edwm} (anomalous transport). In EDWM, saturated wave modes and resulting transport coefficients are predicted from linear growth rates, frequencies and other plasma parameters using quasi-linear theory. Only ion scale turbulence is considered in EDWM, such as ion temperature gradient and trapped electron modes. The simulations consider corrections to turbulence from collisionality and $E \times B$ shear. Five modes are considered in the poloidal mode spectrum, namely $k_\perp\rho_\mathrm{H} = \{0.15, 0.2, \sqrt{0.1}, 0.4, 0.5\}$, where $k_\perp$ is the poloidal wave number, and $\rho_\mathrm{H} = c_\mathrm{s,H}/\Omega_\mathrm{c,H}$ is the hydrogen gyro-radius ($c_\mathrm{s,H}$ is the thermal sound speed, and $\Omega_\mathrm{c,H}$ is the hydrogen gyro-frequency). A Casati-filter~\cite{casati} for rescaling of the flux contributions from each mode is applied, similar to what has been implemented in the quasi-linear transport model QuaLiKiz~\cite{qlk}. A Bohm semi-empirical anomalous heat diffusivity~\cite{bgb} has been added on top of the EDWM and NCLASS diffusivities with a correction factor of 0.1,
\begin{equation}
	\chi_\mathrm{i/e} = \chi_\mathrm{i/e,NCLASS} + (\chi_\mathrm{i/e,EDWM} + 0.1\chi_\mathrm{i/e,Bohm})f_\mathrm{ETB},
\end{equation}
where $f_\mathrm{ETB}$ is the ETB correction during H-mode confinement. This additional term is introduced to enhance transport in the edge/pedestal region, where EDWM frequently underestimates heat diffusivities. The specific correction factor of 0.1 matches simulations of JET experiments with EDWM + Bohm predictive transport.

The particle diffusivity in the scrape-off layer is determined from the computed diffusivity at the last closed flux surface $D_\mathrm{sep}$ (including edge transport barrier in H-mode confinement). A diffusivity $D_\mathrm{bnd}$ is set at the EDGE2D boundary surface. The full diffusivity profile then follows
\begin{equation}\label{eq:difsol}
D(x) = \cases{D_\mathrm{bnd}+(D_\mathrm{sep}-D_\mathrm{bnd})Y_\mathrm{SOL}^{x/\Delta_\mathrm{SOL}}&for $D_\mathrm{sep} \geq D_\mathrm{bnd}$,\\D_\mathrm{ETB}(x)& for $D_\mathrm{sep} < D_\mathrm{bnd}$,}
\end{equation}
\begin{equation}\label{eq:difsol2}
D_\mathrm{ETB}(x) = \cases{D_\mathrm{sep} & for $x < \Delta_\mathrm{ETB}$, \\
D_\mathrm{tanh}(x) & for $\Delta_\mathrm{ETB} \leq x < \Delta_\mathrm{ETB}+\Delta_\mathrm{tanh}$ ,\\
D_\mathrm{bnd} & for $x \geq \Delta_\mathrm{ETB}+\Delta_\mathrm{tanh}$,
}
\end{equation}\begin{equation}
D_\mathrm{tanh}(x) = D_\mathrm{sep} + \frac{D_\mathrm{bnd}-D_\mathrm{sep}}{2}\left[1 + \tanh\left(2\pi\left[\frac{2(x-\Delta_\mathrm{ETB})}{\Delta_\mathrm{tanh}}-1\right]\right)\right],
\end{equation}
where $x$ is the distance from the separatrix along the mid-plane. The model parameters are $D_\mathrm{bnd} = 0.3$\,m$^2$/s, $Y_\mathrm{SOL} = 2.5\times 10^{-3}$, $\Delta_\mathrm{ETB} = 0.5$\,cm, $\Delta_\mathrm{tanh} = 0.3$\,cm, and $\Delta_\mathrm{SOL} = 7.81$\,cm. The thermal diffusivity profiles are defined by the same model function (continuations of $\chi_\mathrm{e,sep}$ and $\chi_\mathrm{i,sep}$, respectively), with the same values of $Y_\mathrm{SOL}$, $\Delta_\mathrm{ETB}$, $\Delta_\mathrm{tanh}$ and $\Delta_\mathrm{SOL}$ as for the particle diffusivity, but with $\chi_\mathrm{e,bnd} = \chi_\mathrm{i,bnd} = 1.0$\,m$^2$/s.

\begin{figure}\centering
\includegraphics[width=.8\textwidth]{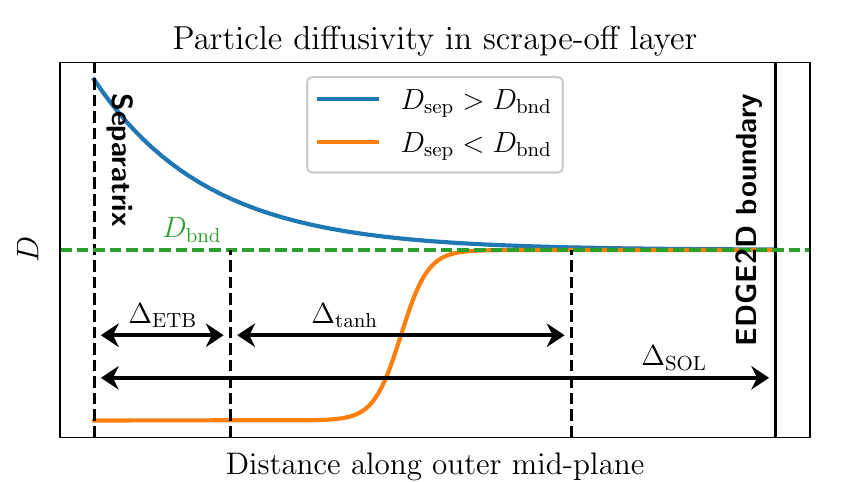}
\caption{Particle diffusivity in the SOL for two different cases of $D_\mathrm{sep}$ (diffusivity at the separatrix calculated by JETTO).}\label{fig:dsol}
\end{figure}

\subsection{Equilibrium \& MHD}

The equilibrium was calculated from predicted pressure and current drive in the core plasma (inside the separatrix) by solving the Grad--Shafranov equation for a fixed separatrix geometry using ESCO~\cite{jetto}. The separatrix geometry was calculated for the 15MA/5.3T baseline scenario~\cite{equil}, using the free-boundary equilibrium solvers CORSICA~\cite{corsica} and DINA~\cite{dina}. It is assumed that the same separatrix shape can be used for the flat-top cases studied in this paper, which all have the same $I_\mathrm{p}/B_\mathrm{tor}$ ratio as the 15MA/5.3T scenario. A Kadomtsev model~\cite{sawt} has been used for sawtooth triggering and relaxation. 

JETTO includes a continuous ELM model, which has been used with the assumption $\alpha \geq \alpha_\mathrm{crit} = 1.8$ for triggering of type-I ELMs. Ideal MHD calculation of similar scenarios have predicted slightly higher values for $\alpha_\mathrm{crit}$ between 2.0 and 2.5~\cite{acrit}. The way that the continuous ELM model in JETTO gradually increases transport when $\alpha > \alpha_\mathrm{crit}$ means that $\alpha$ saturates at a level slightly above $\alpha_\mathrm{crit}$. For this reason, a value $\alpha_\mathrm{crit} = 1.8 < 2.0$ has been selected. H-mode operation with $\alpha < \alpha_\mathrm{crit}$ corresponds more closely to an ELM-free or type-III ELM regime, with no up-scaling of the edge transport barrier. The impact of discrete ELMs on power loads and sputtering from the divertor is not taken into account in these simulations. The presented scenarios are designed for the commissioning of the ELM mitigation systems, and the efficiency of the mitigation systems at the time of running the scenarios is uncertain. 

\subsection{Summary of scenarios}\label{sec:scen}

A total of 6 cases have been modelled with JINTRAC, labelled as follows:
\begin{itemize}
\item Case A: 5MA/1.8T hydrogen plasma with no added helium and 30\,MW of ECRH.
\item Case B: 7.5MA/2.65T H-plasma with no added helium, and 20\,MW of ECRH + 33\,MW of NBI.
\item Case C.1: 7.5MA/2.65T H-plasma with no added helium, and 30\,MW of ECRH + 33\,MW of NBI.
\item Case C.2: Repetition of part of the case C.1, but with a smaller amount of neon being gas puffed for assisting the beam stopping.
\item Case D: 7.5MA/2.65T H-plasma with $\sim$10\,\% added helium, and 20\,MW of ECRH + 33\,MW of NBI.
\item Case E: 7.5MA/2.65T H-plasma with $\sim$10\,\% added helium, and 30\,MW of ECRH + 33\,MW of NBI.
\end{itemize}

\section{Results}

\subsection{5MA/1.8T scenario}

\begin{figure}\centering
\includegraphics[width=.99\textwidth]{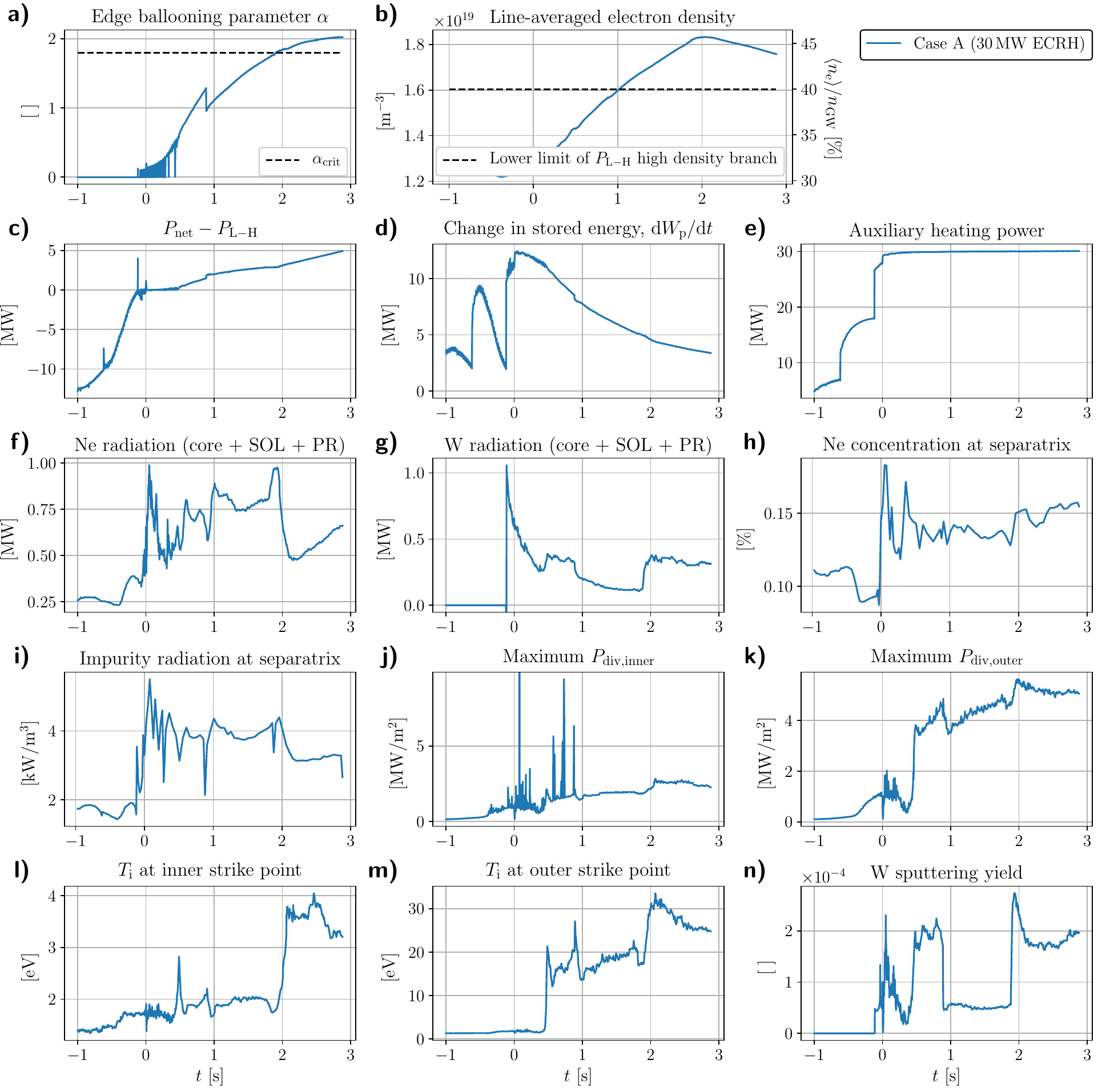}
\caption{ITER 5MA/1.8T hydrogen plasma. The horizontal axis in each of the plots is time [s], offset such that $t = 0$ corresponds to the time of full auxiliary power injection. Auxiliary heating is done by ECRH only.}\label{5MAMisc}
\end{figure}

\begin{figure}\centering
\includegraphics[width=.99\textwidth]{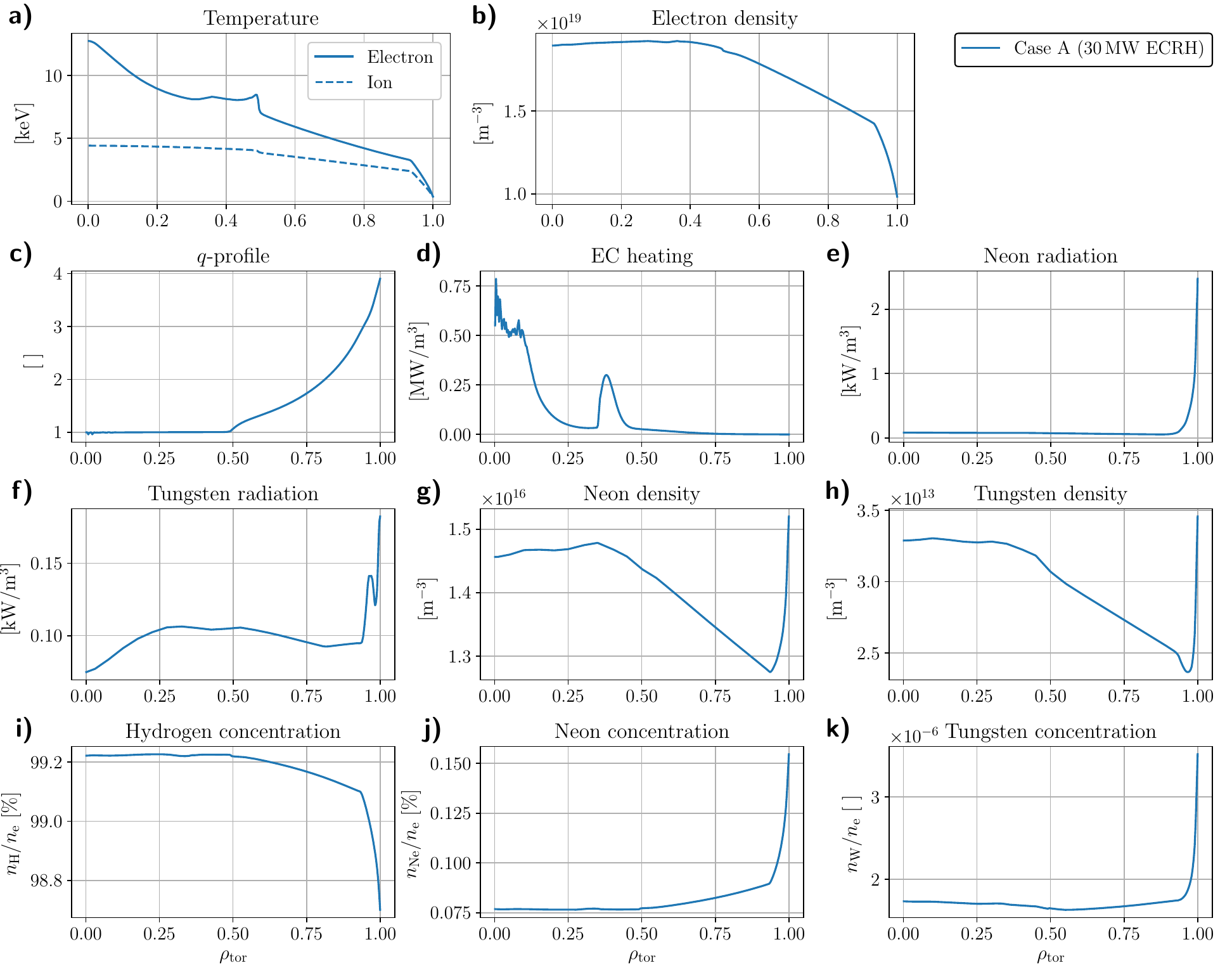}
\caption{Core profile data for case A at the end time of the simulation ($t = 2.88$\,s).}\label{5MAMP}
\end{figure}

The 5MA/1.8T scenario is planned for the PFPO-1 stage of operation, where NBI is unavailable~\cite{loarte}. Consequently, auxiliary heating is done by microwave frequency only. Case A in Figure~\ref{5MAMisc} was run with 30\,MW ECRH. As demonstrated in Figure~\ref{5MAMisc}.a, the edge ballooning parameter exceeds $\alpha_\mathrm{crit}$. The net power difference above the L--H threshold is still increasing towards the end of the simulation, indicating that type-I ELMy H-mode can be maintained for long-pulse operation. In the early phases of H-mode operation, before $t \approx 1$\,s, the line averaged electron density is below $n_{\mathrm{e},\mathrm{min}}$. The Martin08 $P_\mathrm{L-H}$ scaling law has been assumed also in this regime, which is likely an underestimation of the power threshold. For this reason, the transition to H-mode might occur later in practice than what the modelling shows. As long as the plasma can be sufficiently fuelled to reach the high-density branch of $P_\mathrm{L-H}$, the modelling assumptions do not change the conclusion that the scenario can operate in stable ELMy H-mode.

The results of the presented 30\,MW case indicate that a corresponding 20\,MW case would not be able to operate in H-mode. This can be understood from comparing Figures~\ref{5MAMisc}.c with \ref{5MAMisc}.d. At the end of the simulation, $P_\mathrm{net}$ is about 4.9\,MW above $P_\mathrm{L-H}$. An additional margin of 3.4\,MW ($\mathrm{d}W_\mathrm{p}/\mathrm{d}t$ at the end of the simulation in Figure~\ref{5MAMisc}.d) can be expected in steady-state conditions, where $\mathrm{d}W_\mathrm{p}/\mathrm{d}t \rightarrow 0$, totalling to $P_\mathrm{net} - P_\mathrm{L-H} \approx 8.3$\,MW. Dropping auxiliary power by 10\,MW in these conditions would then place the net power about 1.7\,MW below the power threshold. However, it should be noted that the density operates at a level above the lower limit of the high-density branch, as seen in Figure~\ref{5MAMisc}.b. Assuming operation exactly at $\langle n_\mathrm{e}\rangle = n_\mathrm{e,min} \approx 0.4 n_\mathrm{GW}$, eq.~(\ref{eq:plh}) predicts a power threshold $P_\mathrm{L-H} \approx 20.1$\,MW. The estimated core radiation losses exceed the ohmic heating by almost 0.3\,MW. Subtracting this from 20\,MW of auxiliary power results in a net power of 19.7\,MW during steady-state conditions, which is 0.4\,MW less than $P_\mathrm{L-H}$.

The benefit of helium seeding has not yet been explored for this scenario. Assuming a 15\,\% (i.e.\ about 3\,MW) reduction of $P_\mathrm{L-H}$ by helium seeding, keeping the same assumptions about the ohmic heating and impurity radiation as demonstrated in the presented case in Figure~\ref{5MAMisc}, would bring the net power about 2.6\,MW above the power threshold during 20\,MW ECRH operation. Comparing Figures.~\ref{5MAMisc}.a and \ref{5MAMisc}.c, this is very close to the power difference at which $\alpha$ exceeds $\alpha_\mathrm{crit}$ in the demonstrated case. However, taking into account uncertainties in the underlying power threshold assumptions, it is not evident that an ELMy H-mode operational space exists in the helium seeded 5MA/1.8T hydrogen plasma scenario at 20\,MW ECRH.

Since there is no NBI heating for this scenario, neon is not required for beam stopping. However, neon puffing is still used for suppressing the divertor power loads. As can be seen in Figure~\ref{5MAMisc}.h, the neon concentration at the separatrix finishes around 0.15\,\%, which is significantly lower than the corresponding concentration levels for the NBI scenarios (see Figures~\ref{HMisc}.k and \ref{HeMisc}.k), which is of the order of a few percent. Reduced lifetime of the divertor would be expected for long-pulse operation with power loads above about 10\,MW/m$^2$~\cite{IRP}. However, the divertor power loads never exceed 5\,MW/m$^2$, besides single bursts on the inner divertor target plates during the initial stage of the H-mode, as seen in Figures~\ref{5MAMisc}.j and \ref{5MAMisc}.k.

Tungsten sputtering from the divertor was not included during the L-mode phase of the simulation, as can be seen in Figure~\ref{5MAMisc}.g. It was included at the start of the dithering phase, at $t = -0.12$\,s. During the H-mode phase, there were two instances of rapid increase of the strike point ion temperature and the tungsten sputtering yield from the divertor, at $t \approx 0.45$\,s and $t = 1.88$\,s, as seen in Figures~\ref{5MAMisc}.l, \ref{5MAMisc}.m and \ref{5MAMisc}.n. However, the associated tungsten radiation in the plasma volume remained at modest levels, below 400\,kW, thoughout most of the H-mode phase. 

Figure \ref{5MAMP} shows miscellaneous core 1D profile data for the 5MA/1.8T scenario at the end time of the simulations ($t = 2.88$\,s). The additional 10\,MW on the upper EC launcher gives no power absorption inside $\rho_\mathrm{tor} = 0.35$, with a peak resonance around $\rho_\mathrm{tor} = 0.38$, as seen in Figure~\ref{5MAMP}.d. The non-monotonic electron temperature profile of Figure~\ref{5MAMP}.a is a result of both the off-axis ECRH and sawtooth crashes up to $\rho_\mathrm{tor} \approx 0.5$ (domain with $q \approx 1$ in Figure~\ref{5MAMP}.c). There is some degree of impurity accumulation, as seen in the peaked impurity density profiles of Figures~\ref{5MAMP}.g and \ref{5MAMP}.h. However, core impurity concentrations are relatively low, with less than 0.1\,\% of neon, and less than 2$\times 10^{-6}$ of tungsten, as seen in Figures~\ref{5MAMP}.j and \ref{5MAMP}.k, respectively. The higher temperatures of the deep core also suppresses impurity radiation in these regions, with radiation being the highest in the edge (see Figures~\ref{5MAMP}.e and \ref{5MAMP}.f). Figure \ref{5MAMisc}.i shows the time trace of the total radiation density at the separatrix. Comparing these values to those of the other presented scenarios (Figures~\ref{HMisc}.l and \ref{HeMisc}.l), which has higher neon seeding in order to reduce NBI shine-through, the radiation density is an order of magnitude lower for the 5MA/1.8T scenario.

\subsection{7.5MA/2.65T scenario without helium}

\begin{figure}\centering
\includegraphics[width=.99\textwidth]{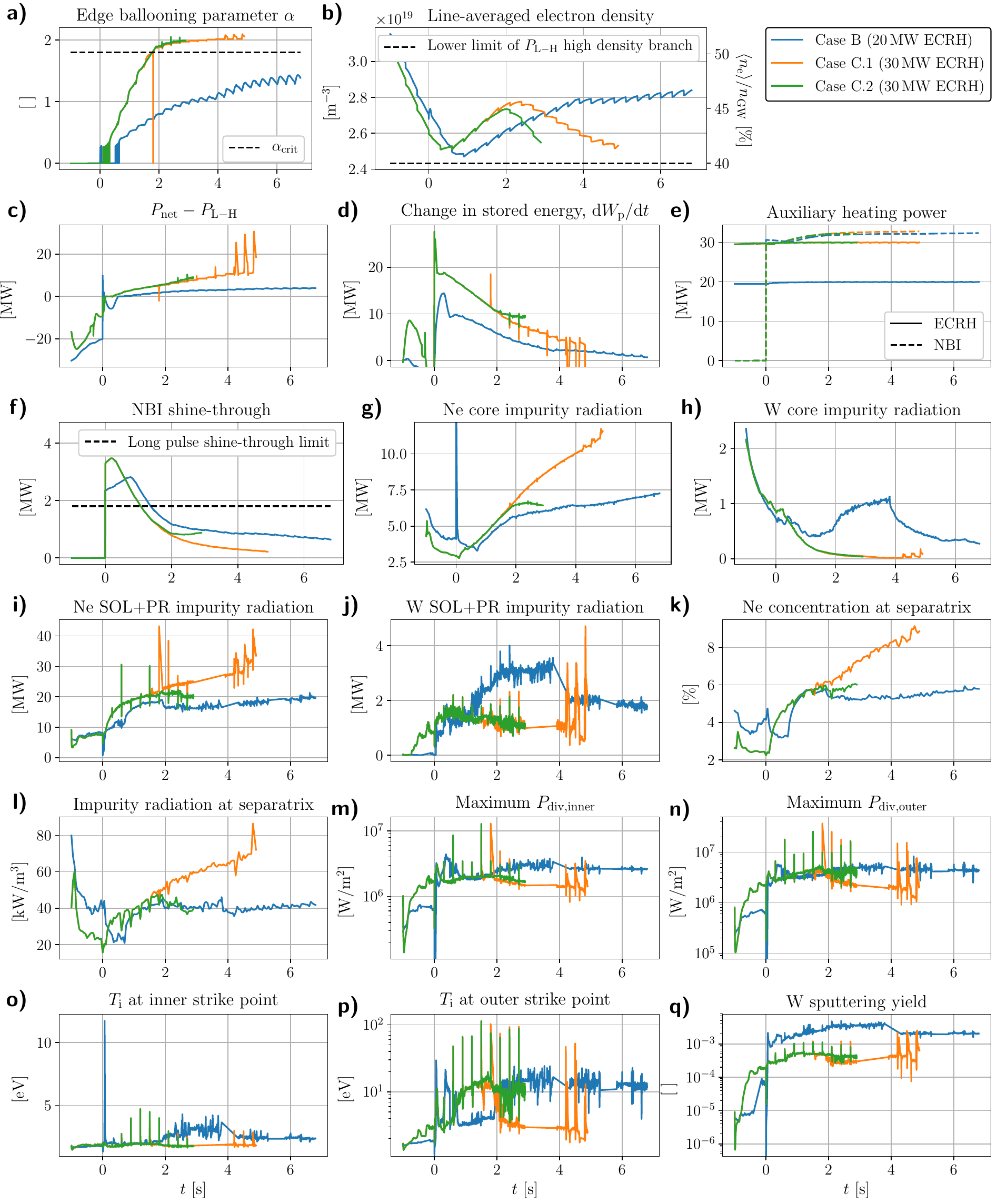}
\caption{ITER 7.5MA/2.65T hydrogen plasma simulations without a helium minority. The horizontal axis in each of the plots is time [s], offset such that $t = 0$ corresponds to the time of full auxiliary power injection.}\label{HMisc}
\end{figure}

\begin{figure}\centering
\includegraphics[width=.99\textwidth]{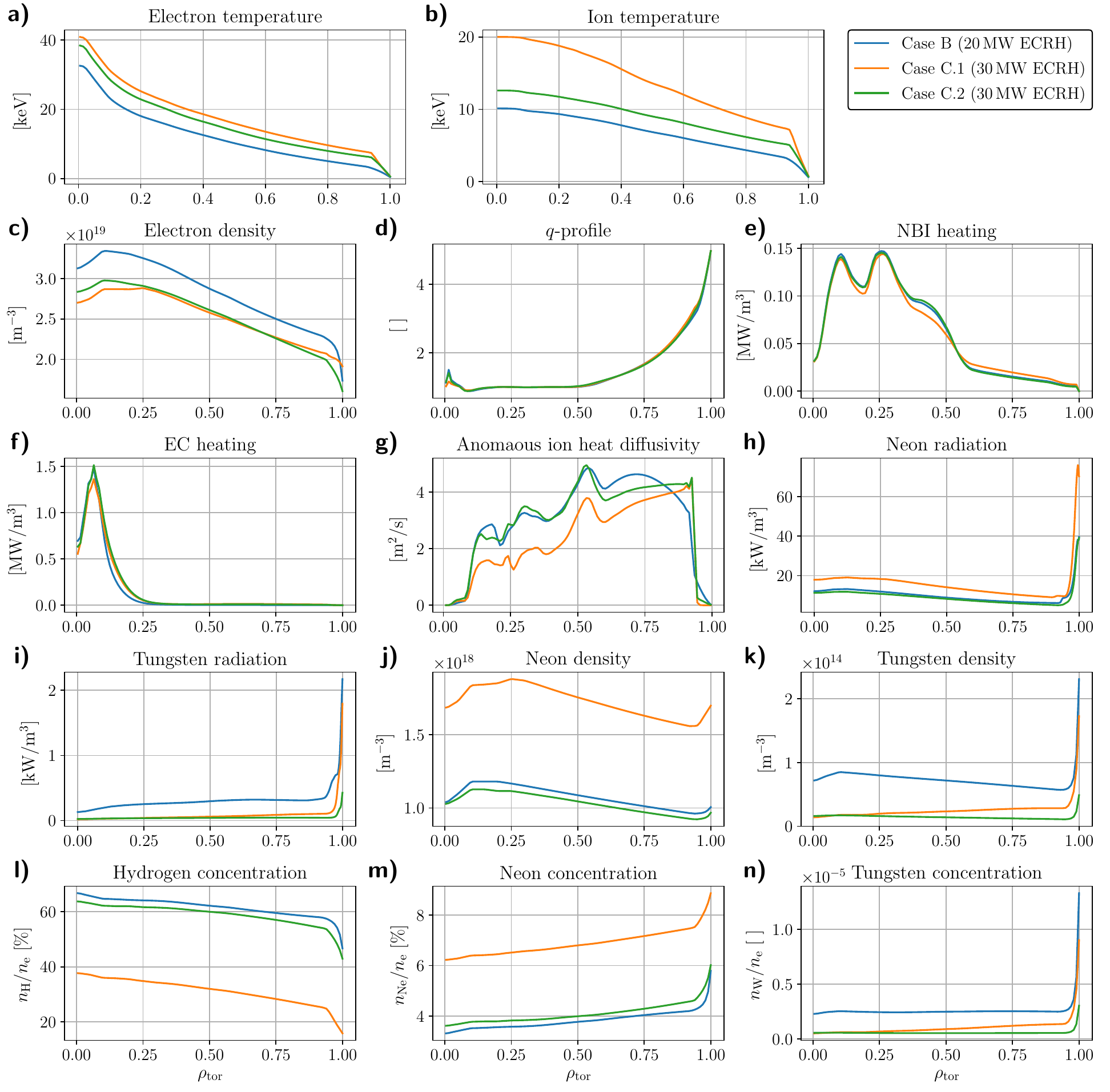}
\caption{Core profile data for cases B, C.1 and C.2 at the end time of the simulations (case B: $t = 6.80$\,s, case C.1: $t = 4.90$\,s, case C.2: $t = 2.90$\,s).}\label{HMP}
\end{figure}

The 7.5MA/2.65T hydrogen plasma scenario is planned for PFPO-2, including heating and current drive from both EC and NBI. The result of the JINTRAC simulations of cases B, C.1 and C.2, which do not include any helium, are presented in Figures~\ref{HMisc} and \ref{HMP}. It is apparent from Figure~\ref{HMisc}.a that case B, accessing H-mode with 20\,MW of ECRH, is likely not to be able to reach type-I ELMy H-mode, with the edge ballooning parameter $\alpha$ remaining below the estimated $\alpha_\mathrm{crit}$. Both cases C.1 and C.2 have $\alpha$-parameters exceeding $\alpha_\mathrm{crit}$, indicating that 30\,MW of ECRH is required for stable ELMy H-mode operation of the hydrogen plasma without a helium minority. It should be noted that the line-averaged electron density of case B (see Figure~\ref{HMisc}.b) stabilizes about 14 -- 16\,\% above the lower limit of the high-density branch $n_{\mathrm{e}, \mathrm{min}}$, meaning that it is theoretically possible to reach $11\,\% \approx 4$\,MW lower $P_\mathrm{L-H}$ when operating exactly at $\langle n_\mathrm{e}\rangle = n_{\mathrm{e}, \mathrm{min}}$. According to Figure~\ref{HMisc}.c, case B stabilizes at around 4\,MW net power above the L-H power threshold, meaning that the margin above the threshold could in principle be doubled, which might be sufficient to operate at a stable ELMy H-mode. However, considering the narrow operation margins of case B, combined with the uncertainty of $n_{\mathrm{e}, \mathrm{min}}$ in ITER, access to ELMy H-mode at 20\,MW ECRH cannot be taken for granted.

Case C.1 was excessively seeded with neon to reduce the NBI shine-through, with the final shine-through power around 0.2\,MW (see Figure~\ref{HMisc}.f), well below the estimated long-pulse shine-through limit at 1.8\,MW, and the neon core impurity radiation finishing close to 12\,MW (see Figure~\ref{HMisc}.g). This should be compared to case C.2, which is the same case as C.1, but with reduced neon seeding. The NBI shine-through for this case stabilizes around 0.9\,MW, and the neon core impurity radiation at around 6.5\,MW, meaning that there is a net gain of about 4.8\,MW input power compared to case C.1. However, this relative gain is cancelled in $P_\mathrm{net}$ to some extent by differences in $\mathrm{d}W_\mathrm{p}/\mathrm{d}t$. The net power $P_\mathrm{net}$, which is compared against $P_\mathrm{L-H}$ in Figure~\ref{HMisc}.c, is calculated as $P_\mathrm{aux} + P_\mathrm{ohm} - P_\mathrm{rad} - \langle\mathrm{d}W_\mathrm{p}/\mathrm{d}t\rangle$, where $W_\mathrm{p}$ is the total stored energy. Neither case C.1 nor C.2 have reached stationary conditions, which would have been indicated by $\mathrm{d}W_\mathrm{p}/\mathrm{d}t \approx 0$ in Figure~\ref{HMisc}.d. However, the simulations have run sufficiently long to conclude that type-I ELMy H-mode is accesible in the 30\,MW ECRH scenario, as $\alpha$ has already stabilized above $\alpha_\mathrm{crit}$, and a reduction of $\langle\mathrm{d}W_\mathrm{p}/\mathrm{d}t\rangle$ will only increase the margin above the L-H power threshold.

\begin{figure}\centering
\includegraphics[width=.99\textwidth]{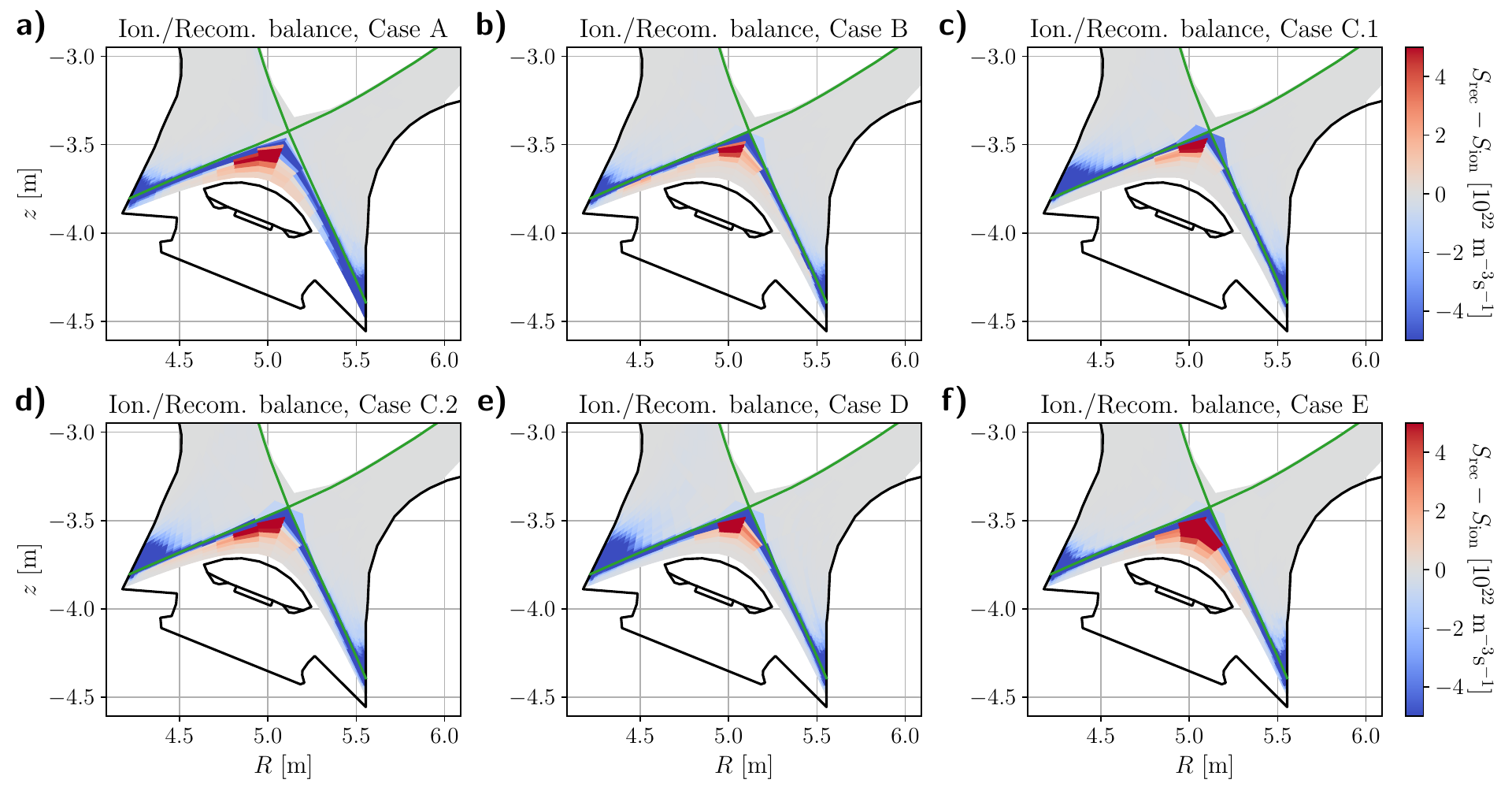}
\caption{Difference between recombination and ionisation rates for all six cases. The colourmaps have been truncated at $\pm 5\times 10^{22}$\,m$^{-3}$s$^{-1}$.}\label{ionRec}
\end{figure}

As mentioned in Section~\ref{sec:fuel}, there is a risk of reaching regimes of full detachment if the edge region is cooled down significantly by gas puffing and impurity radiation. To verify that neither of the presented scenarios has reached full detachment, including Case C.1, which was excessively seeded by neon, the recombination rates are compared against the ionisation rates for all cases in Figure~\ref{ionRec}. Full detachment would be indicated by a region around the separatrix, in particular around the x-point, where $S_\mathrm{rec} > S_\mathrm{ion}$. All of the cases have a limited domain in the private region where $S_\mathrm{rec} > S_\mathrm{ion}$. This means that the modelling indicates that the scenarios are only partially detached at most.

In Figure~\ref{HMisc}.k, the neon concentration is calculated as a fraction of the electron density. This value stabilized around 6\,\% for cases B and C.2, whereas it reached 9\,\% (and increasing) towards the end of case C.1, indicating an almost pure neon plasma at the separatrix for that case (the full neon concentration profile at the end time of each simulation can be seen in Figure~\ref{HMP}.m). Although neon is more efficient at sputtering tungsten from the divertor than hydrogen, case C.1 did not show higher tungsten radiation compared to cases B and C.2 (see Figures~\ref{HMisc}.h and \ref{HMisc}.j). This could be a consequence of the high neon radiation in the SOL (Figure~\ref{HMisc}.i), which keeps the strike point ion temperature down and reduces the overall sputtering from the divertor. Case B demonstrated relatively high strike point ion temperatures (Figures~\ref{HMisc}.o and \ref{HMisc}.p) and tungsten sputtering yield (Figure~\ref{HMisc}.q). However, with tungsten core impurity radiation below the MW range, the tungsten content in the plasma volume is not sufficient to substantially impact the H-mode operation. The maximum power loads on the inner and outer divertor target plates stabilize at a few MW/m$^2$ (see Figures~\ref{HMisc}.m and \ref{HMisc}.n), besides a couple of discrete bursts, particularly on the outer target plates in cases C.1 and C.2. The bursts in divertor power loads are associated with particle and heat fluxes in the outer part of the plasma due to large predicted anomalous transport following sawtooth crashes.

Core profile data for cases B, C.1 and C.2 is shown in Figure~\ref{HMP}. Case C.1 has significantly higher ion temperature (Figure~\ref{HMP}.b), which could be explained by the lower anomalous ion diffusivity (Figure~\ref{HMP}.g). Neither of the cases show the same non-monotonicity of the electron temperature as case A. This is because the additional 10\,MW ECRH is injected from a hypothetical equatorial launcher in O-mode, allowing for more on-axis electron heating. There is a non-monotonic deposition of the NBI power density, but the relatively spread out absorption in $\rho_\mathrm{tor}$ does not affect the shape of the electron or ion temperature profiles significantly. Again, sawtooth crashes extend to $\rho_\mathrm{tor} \approx 0.5$, as seen in Figure~\ref{HMP}.d. While impurity radiation profiles for cases B, C.1 and C.2 are qualitatively the same as the ones for case A, they are between 1 and 2 orders of magnitude larger (compare Figures~\ref{HMisc}.h and \ref{HMisc}.i with Figures~\ref{5MAMisc}.e and \ref{5MAMisc}.f, respectively). This is due to the additional requirement for neon to reduce the NBI shine-through, which in turn sputters tungsten from the divertor more effectively than hydrogen. The hydrogen fraction is significantly lower in cases B, C.1 and C.2 compared to case A (compare Figures~\ref{HMP}.l with \ref{5MAMP}.i), and case C.1 in particular, where it is as low as 16\,\% at the separatrix. This is again due to the higher neon content of these cases.

\subsection{7.5MA/2.65T scenario seeded with helium}

\begin{figure}\centering
\includegraphics[width=.99\textwidth]{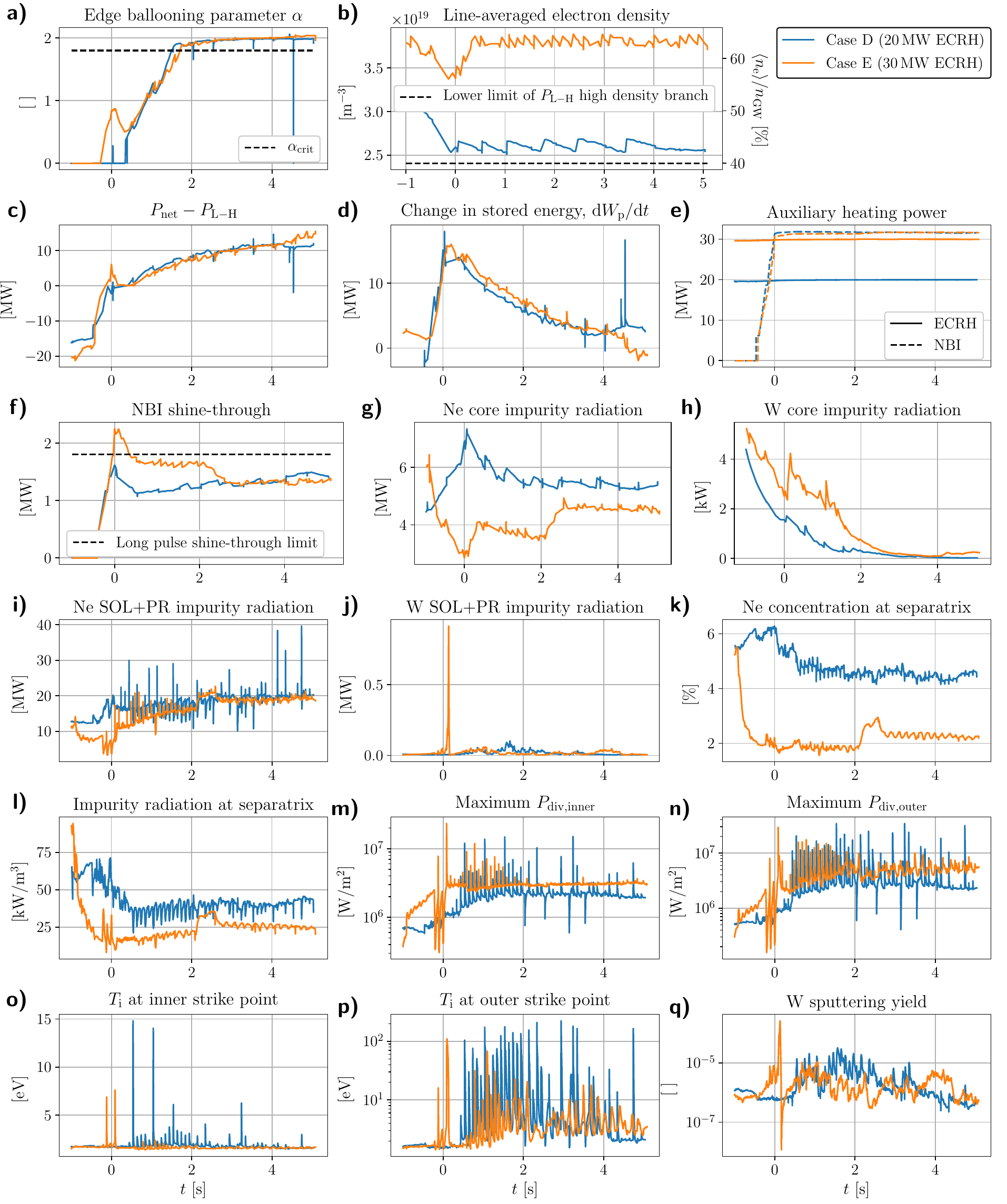}
\caption{ITER 7.5MA/2.65T hydrogen plasma simulation with a helium minority. The horizontal axis in each of the plots is time [s], offset such that $t = 0$ corresponds to the time of full auxiliary power injection.}\label{HeMisc}
\end{figure}

\begin{figure}\centering
\includegraphics[width=.99\textwidth]{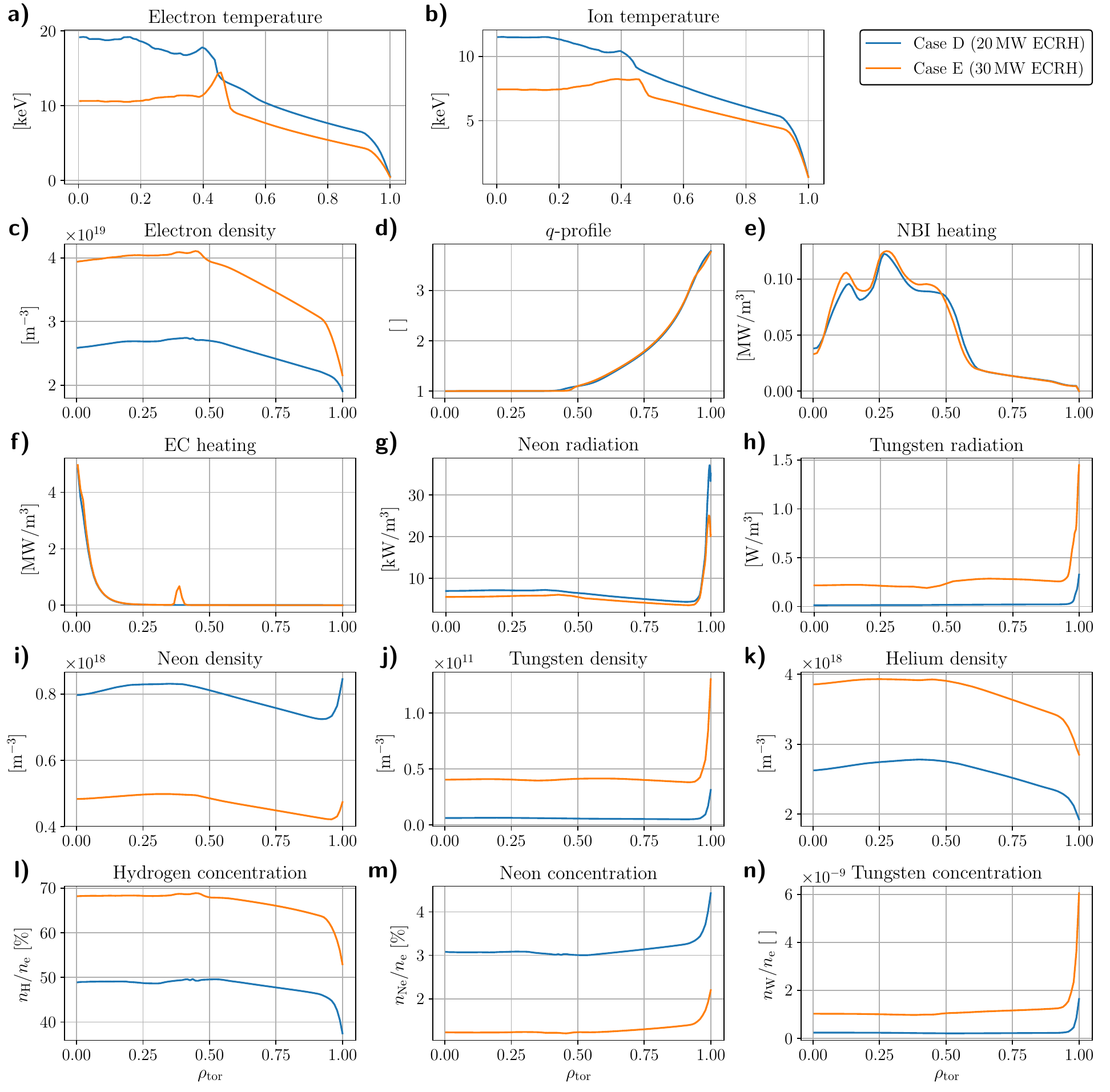}
\caption{Core profile data for cases D and E at the end time of the simulations (case D: $t = 5.04$\,s, case E: $t = 5.08$\,s).}\label{HeMP}
\end{figure}

In the two cases presented in Figure~\ref{HeMisc} (cases D and E), the 7.5MA/2.65T hydrogen plasma has been seeded with helium for presumed lowering of the L--H power threshold, as discussed in Section~\ref{sec:PLH}. This increases the net power above the threshold for improved H-mode quality. Figure \ref{HeMisc}.a confirms that stable type-I ELMy H-mode operation is reached both for the 20\,MW and the 30\,MW ECRH cases. Unlike cases A, B, C.1 and C.2, the cases D and E use a combination of hydrogen gas puffing and pellet injection to fuel the plasma, as discussed in Sec.~\ref{sec:fuel}. The pellet injection rate is adapted for reaching target averaged electron densities. The target density for the 20\,MW ECRH case (case D) is set close to $n_{\mathrm{e}, \mathrm{min}}$, whereas the target density for the 30\,MW ECRH case (case E) is between 55 and 60\,\% above $n_{\mathrm{e}, \mathrm{min}}$, corresponding to $\langle n_\mathrm{e}\rangle/n_\mathrm{GW}$ between 62 and 64\,\%, as seen in Figure~\ref{HeMisc}.b. The higher target density for case E is chosen to demonstrate a wider operational space with 30\,MW ECRH, although it increases the L--H power threshold by up to about 40\,\% = 12\,MW.

The NBI power is gradually increased until it finally reaches its full power at $t = 0$, as demonstrated in Figure~\ref{HeMisc}.e. The NBI shine-through power stabilizes around 1.3 -- 1.5\,MW for both cases (Figure~\ref{HeMisc}.f), which is below the assumed limit of 1.8\,MW for long-pulse operation. Slightly lower neon content is required for sufficient beam stopping in case E compared to case D because of the higher averaged density. This is also reflected in the lower neon core radiation of case E in Figure~\ref{HeMisc}.g. However, the neon radiation in the SOL and private region is similar for the two cases, stabilising around 20\,MW, as seen in Figure~\ref{HeMisc}.i. As for the 7.5MA/2.65T cases with no helium (cases B, C.1 and C.2), the tungsten radiation is negligible compared to the neon radiation, as seen in Figures~\ref{HeMisc}.h and \ref{HeMisc}.j.

The maximum power load on the divertor target plates stabilizes around 2 -- 6\,MW/m$^2$ (see Figures~\ref{HeMisc}.m and \ref{HeMisc}.n), which is well within acceptable long-pulse operational limits. However, there are frequent bursts in the power loads, in particular at the outer target during the initial stages of the H-mode. Again, the largest bursts coincide with wide sawtooth crashes, inducing significant particle and heat fluxes in the outer core plasma via bursts in anomalous diffusivity. Since the bursts are very short and drop in frequency during later stages of the H-mode confinement, they are unlikely to cause significant reduction of the divertor lifetime. The ion temperature of the strike points are also sustained at relatively low levels at the order of 1\,eV, with the exception of discrete bursts, as seen in Figures~\ref{HeMisc}.o and \ref{HeMisc}.p. This causes a very low tungsten sputtering yield of the order 10$^{-6}$ (see Figure~\ref{HeMisc}.q), which is 2 -- 3 orders of magnitude lower than any of the other presented cases.

Similarly to case A, cases D and E show some degree of non-monotonicity of the temperature profiles (Figures~\ref{HeMP}.a and \ref{HeMP}.b), due to a combination of sawtooth crashes and off-axis heating. Cases D and E used a magnetic reconnection factor of 1.0 for the Kadomtsev model, whereas the other cases used a reconnection factor of 0.3. The main reason for the higher reconnection factor was to make a pessimistic assumption about the impact of sawteeth on the scenarios, such as higher amplitude oscillations of the plasma state and their impact on e.g.\ tungsten sputtering. The lower reconnection factor is more in agreement with experimental results from JET. The higher reconnection factor of Cases D and E could cause steeper pressure gradients across the $q = 1$ boundary close to $\rho_\mathrm{tor} = 0.5$ (see $q \approx 1$ domain in Figure~\ref{HeMP}.d). With neon gas puffing being overall lower in cases D and E compared to the corresponding cases with no helium (cases B, C.1 and C.2), both the neon density (Figure~\ref{HeMP}.i), concentration (Figure~\ref{HeMP}.m) and radiation density (Figure~\ref{HeMP}.g) are lower (compare against Figures~\ref{HMP}.j, \ref{HMP}.m and \ref{HMP}.h, respectively). The tungsten density and radiation levels are exceptionally low for cases D and E compared to other cases, with densities of the order 10$^{10}$\,m$^{-3}$ and radiation densities below 1\,W/m$^3$ in most of the core plasma, which is a result from the low sputtering yield. However, with the neon concentration being of the order of a few \% in the core plasma, it is enough to significantly reduce the hydrogen concentration, as seen in Figure~\ref{HeMP}.l. Case D, which operates at lower density than case E, and consequently requires more neon for sufficient beam stopping, has a core concentration of hydrogen of less than 50\,\%.

\section{Conclusions and discussion}

ITER will undergo a Pre-Fusion Power Operation (PFPO) phase, during which stable H-mode confinement will be demonstrated and several operational systems will be commissioned. For the commissioning of ELM mitigation systems, stable operation in type-I ELMy H-mode is required. The presented paper investigates the possibility to operate the 7.5MA/2.65T and 5MA/1.8T hydrogen plasma scenarios for this purpose, using core plasma and edge/SOL/divertor integrated modelling tools. Furthermore, it is tested whether the currently planned ECRH power capacity is sufficient for stable ELMy H-mode operation or a proposed 10\,MW upgrade of the EC systems is required. The L--H power threshold of the hydrogen plasma is expected to be a factor $\sim$2 larger compared to equivalent deuterium plasma scenarios. Operation at low density and seeding the plasma with a helium minority are both methods used to attempt to reduce the power threshold for easier H-mode access. While H-mode access and margin to the L--H transition can be estimated with core plasma modelling, a complete assessment of the viability of these scenarios in ITER requires full core/edge/SOL modelling, which is done for the study presented in this paper. Many of the strategies followed to lower the L--H transition have direct implications on the SOL and plasma--wall interaction (PWI) aspects with effects (such as full detachment or excessive W contamination of the core plasma) that can counter balance the expected effects from core plasma simulations and make, in practice, the scenarios for H-mode access and sustainment in ITER not viable. The modelling performed in this paper identifies which of such strategies will lead to realizable PFPO H-mode scenarios with core-edge-PWI factors taken into account.

The 5MA/1.8T scenario is planned for the early stages of PFPO, during which only electron cyclotron resonance power systems are available for auxiliary heating, whereas the 7.5MA/2.65T scenario is also assisted by NBI heating. Neon is seeded by gas puffing to reduce NBI shine-through and to prevent excessive power loads to the divertor. Excessive neon seeding can potentially impede access to type-I ELMy H-mode by reducing the margin of net power above the L--H power threshold via core impurity radiation. The presented modelling, which includes self-consistent treatment of neon seeding, tungsten sputtering from the divertor, impurity transport, and atomic physics, is used to investigate whether the neon seeding can be optimized for sufficient beam stopping and divertor cooling without compromising ELMy H-mode access for the different scenarios. All modelled cases start from L-mode confinement at reduced auxiliary power, eventually ramping up to full input power to demonstrate a stable L--H transition. Once a quasi-stationary H-mode flat-top has been reached, the edge ballooning parameter $\alpha$ is compared against the estimated threshold, $\alpha_\mathrm{crit}$, for triggering type-I ELMs (using a continuous ELM model in this paper).

Modelling results indicated that the 7.5MA/2.65T hydrogen plasma scenario might more readily access ELMy H-mode either through seeding by helium or a 10\,MW upgrade of the EC systems. The success of the helium seeded scenario without the ECRH power upgrade relies on the assumptions that a) the lower limit of the high-density branch of the L--H power threshold is similar to or lower than 0.4$n_\mathrm{GW}$, and b) a 10\,\% helium minority reduces $P_\mathrm{L-H}$ by 15\,\%. Both assumptions are based on results from JET experiments, for which it is unknown how well they extrapolate to ITER operation regimes. In particular, the lower limit of the $P_\mathrm{L-H}$ high-density branch has shown to vary significantly with the strike-point configuration at JET. The 10\,MW ECRH upgrade allows for operation in ELMy H-mode at higher density, which has been demonstrated for the helium seeded case. This relaxes the assumption that the lower limit of the $P_\mathrm{L-H}$ high-density branch $n_\mathrm{e,min}$ is at $0.4 n_\mathrm{GW}$, giving more confidence in robust access to H-mode based on the assumed $P_\mathrm{L-H}$ scaling.

The 5MA/1.8T hydrogen plasma scenario has a lower Greenwald density compared to the 7.5MA/2.65T scenario due to the lower plasma current. The scenario can consequently operate at a lower absolute density while staying above the lower limit of the $P_\mathrm{L-H}$ high-density branch. However, the lower available auxiliary power means that the 10\,MW ECRH upgrade is still likely required to operate in a stable ELMy H-mode. With an expected L--H power threshold at around 20\,MW at $\langle n_e\rangle = n_\mathrm{e,min} \approx 0.4 n_\mathrm{GW}$, H-mode confinement can at best be marginally reached with 20\,MW of ECRH, not allowing access to type-I ELMy H-mode operation. 

\section*{Acknowledgements}
JINTRAC was used under licence agreement between Euratom and CCFE, Ref. Ares(2014)3576010 -28/10/2014. This work was part funded by the RCUK Energy Programme [grant number EP/T012250/1], EPSRC Energy Programme [grant number EP/W006839/1], and by ITER Task Agreement C19TD53FE implemented by Fusion for Energy under Grant GRT-869 and Contract OPE-1057. The views and opinions expressed herein do not necessarily reflect those of the ITER Organization. To obtain further information on the data and models underlying this paper please contact PublicationsManager@ukaea.uk.

\section*{References}
\bibliographystyle{iopart-num}
\bibliography{ITER_PFPO_Tholerus_etal_2024_arXiv}

\end{document}